# *Optimizing Storytelling, Improving Audience Retention, and Reducing Waste in the Entertainment Industry*

1st Andrew Cornfeld
School of Data Science
University of Virginia
Charlottesville, VA, USA
cpm6gh@virginia.edu

1st Ashley Miller,
School of Data Science
University of Virginia
Charlottesville, VA, USA
asm2fe@virginia.edu

2nd Mercedes Mora-Figueroa
School of Data Science
University of Virginia
Charlottesville, VA, USA
eqa7yg@virginia.edu

2nd Kurt Samuels
School of Data Science
University of Virginia
Charlottesville, VA, USA
nps3cs@virginia.edu

3rd Anthony Palomba
Darden School of Business
University of Virginia
Charlottesville, VA, USA
palombaa@darden.virginia.edu

*Abstract* — Television networks face considerable uncertainty when predicting episodic viewership, with inaccurate forecasts potentially resulting in costly programming decisions. This study introduces a content-aware machine learning framework for episode-level audience prediction, leveraging natural language processing (NLP) features extracted from over 25,000 television episodes across 219 series. We combine prior viewership data with NLP-derived measures, such as emotional tone, cognitive processing, and dialogue structure, captured across each episode's narrative arc. We evaluate the forecasting accuracy of SARIMAX models, rolling XGBoost regressors, and a nested feature selection ensemble with SHAP interpretability. The framework is evaluated on a diverse set of shows including *Better Call Saul*, *Abbott Elementary*, *CSI: Miami*, *CSI: Crime Scene Investigation*, *The Shield*, and *Rick and Morty* to assess model performance across genres and storytelling formats. Results demonstrate that prior viewership is a strong baseline predictor, but NLP features add meaningful predictive power for certain series. We also introduce a similarity scoring method using Euclidean distance between aggregate dialogue vectors, enabling content-based comparisons between shows. These findings offer a scalable approach for forecasting audience behavior and guiding programming, development, and marketing decisions.

## 1. INTRODUCTION

Television networks routinely face high financial stakes when launching new shows, as the inability to predict success can lead to millions of dollars in losses from canceled seasons or unaired pilots. Traditional forecasting approaches have leaned heavily on lagging indicators such as historical ratings, time slot positioning, and genre-level trends. While useful at a macro level, these models provide limited insight into episode-level dynamics or narrative structures that influence audience engagement in real time. Increasing interest in using generative AI in Hollywood has shifted interest in not just the construction of TV scripts [1], but also creates opportunities for the systematic deconstruction of TV scripts.

This study introduces a twofold contribution to the emerging field of computational media analytics. First, we develop dynamic episode-level forecasting models that integrate natural language processing (NLP) derived features from television scripts, including emotional tone, cognitive complexity, and narrative structure, with prior viewership data. Second, we evaluate the effectiveness of an autoregressive ensemble learning framework [4] to forecast subsequent episode performance, offering a data-driven complement to creative and strategic decision-making within television networks.

By shifting the unit of analysis from the series level to the episode level, this research offers a more granular understanding of how specific linguistic and emotional elements within dialogue influence audience retention over time. Our central aim is to determine whether these content-derived variables add incremental predictive value beyond traditional metrics. In doing so, we contribute to broader conversations in media studies about how storytelling elements shape audience behavior, while also introducing methodological innovations valuable to both scholars and industry practitioners navigating the evolving television landscape.

## 2. RELATED WORK

Early approaches to predicting television viewership primarily relied on historical ratings and schedule-based variables, employing statistical models such as ARIMA and SARIMA to account for trends and seasonality in audience data [2]. Regression analyses that incorporated factors like time slots, lead-in programs, and competing broadcasts enabled moderately accurate short-term forecasts based on past viewing patterns [3]. While these traditional models were effective for established shows with consistent audiences, they often struggled to adapt to shifts in viewer behavior or to predict performance for new content lacking historical benchmarks.

To address these limitations, researchers have increasingly turned to machine learning techniques, which offer greater flexibility in modeling nonlinear relationships and incorporating a wider array of features [5]. A wide range of models, from linear regression and K-nearest neighbors to random forests, XGBoost, and neural networks, have been applied to viewership prediction, with ensemble methods frequently outperforming individual algorithms. By incorporating historical ratings, demographic data, and scheduling variables, techniques such as gradient boosting and ridge regression have achieved greater predictive accuracy in complex viewing environments. In practice, hybrid models that blend statistical and machine learning approaches, such as SARIMAX, Prophet, and XGBoost, have proven effective at reducing forecasting error across diverse television genres and audience segments.

More recently, natural language processing (NLP) has emerged as a novel approach to viewership prediction by incorporating content-based features derived from scripts or subtitles. Studies have demonstrated that linguistic characteristics, such as sentiment, emotional arcs, thematic complexity, and narrative structure, correlate with audience engagement and ratings [6]. For example, researchers have used LIWC-style emotion dictionaries and network text analysis to quantify narrative originality and emotional tone, finding significant associations with subsequent viewership. By integrating these NLP-derived features with historical and contextual data, predictive models gain access to intrinsic narrative signals, offering a promising avenue for understanding and forecasting audience responses beyond traditional metrics.

## 3. PROPOSED METHOD

### A. Data Description

Our data set consists of approximately 25,257 episodes spanning 219 unique television shows. Each episode record contains 144 variables spanning show metadata, narrative emotion scores, linguistic style, cognitive processes, motivational drives, and engineered features.

Key metadata variables include show title, season, episode number, air date, episode length, IMDb rating, genre, network, viewership in millions, and cancellation status.

Narrative and linguistic features were derived from episode sub caption data, scraped from *OpenSubtitles* using an API, with additional show-level metadata (e.g., parent network, number of episodes, season count) sourced from IMDb.

A central focus of the dataset is the set of 43 linguistic and emotion-related NLP-derived scores, computed per three narrative acts per episode, resulting in 129 variables per episode. These variables include measures of emotional tone (e.g., Anger, Joy, Fear), sentiment (Positive, Negative), cognitive processes (Insight, Cause), and narrative style (Analytic, Clout, Authenticity, Tone).

### B. Data Preprocessing

Prior to modeling, we conducted a series of preprocessing steps to ensure data quality, consistency, and suitability for analysis. First, the number of genre categories was reduced from 54 to 14 broader classifications, and categorical variables such as genre and network were one-hot encoded.

We then engineered new temporal features by extracting the month, year, year-month, day of the week, and season from each episode's air date. A combined year-season variable was also created to support seasonal trend analysis. Additionally, we introduced a categorical variable to indicate whether an episode was a season premiere, season finale, or mid-season episode.

To better capture viewership dynamics, we calculated both the percentage change in viewership from the previous episode and a three-episode moving average. The percentage change was explored as a potential alternative target variable to absolute viewership (in millions), while the moving average was included to help account for longer-term trends and smooth out short-term fluctuations.

A key preprocessing step involved the temporal alignment of lagged features: we shifted each episode's explanatory variables so they corresponded to predicting the next episode's viewership (or its derived metrics). This alignment ensured that all information available at episode t was used to forecast outcomes at episode t+1, reflecting a real-world predictive scenario.

## 4. MODELING APPROACHES

### A. SARIMAX Model

Given the cyclical nature of television shows across seasons, we initially explored a seasonal autoregressive integrated moving average (SARIMA) model. However, the data lacked consistent time intervals between episodes within each show, presenting a major limitation for time series modeling. Interpolating missing time steps would have introduced substantial imputation error, so we instead attempted to recover lost temporal information by incorporating an exogenous variable [7].

To account for irregular airing schedules, we engineered a new feature capturing the time elapsed since the previous episode. While this adjustment yielded reasonable viewership estimates for long-running shows, it performed poorly for newer series with few episodes, due to the model's minimum observation requirements for convergence. As a result, this approach proved ill-suited to our broader objective.

Our goal is to use these data to inform creative decisions around show direction and writing. A model that relies primarily on historical fluctuations in viewership fails to provide meaningful insights into narrative or content features. Similarly, while we considered incorporating natural language scores as exogenous variables, doing so risked compounding forecast error, and more importantly, we could not fully justify their exogeneity, as these scores are inherently tied to the very system influencing viewership.

### B. Rolling XGBoost Model

Our second approach uses XGBoost, which takes data solely from a given show and creates a sequentially retrained XGBoost model that predicts episode viewership using cumulative data from all preceding episodes. We chose to use XGBoost due to its efficiency as well as its ability to capture interactions between variables. This model removes temporal data such as air date, season, and episode numbers. This model begins with Season 1 Episode 3 (to have enough data for training) and continues through all episodes of the show.

One limitation of this model is that it requires prior viewership data, making it unsuitable for predicting the performance of unaired shows. During testing, we found that shows with a large number of episodes take significantly longer to process, sometimes several minutes. While we explored ways to reduce runtime, such as adjusting how frequently the model retrains, we ultimately concluded that the resulting loss in accuracy outweighed the potential time savings.

### C. Feature Selection Model

Since our original boosted model showed promising performance, we began exploring ways to enhance it. One major challenge was the high dimensionality introduced by the natural language processing (NLP) scores. To address this, we developed a nested XGBoost regression approach that narrowed the feature space. The first-stage model identified the top 20 features based on the built-in gain importance metric. These selected features were then used to train a second-stage XGBoost model.

In addition to this feature selection step, we engineered new predictors, including the previous episode's viewership and two rolling averages with window sizes of three and five episodes. To further interpret feature influence, we cross-referenced gain-based importance rankings with Shapley Additive Explanations (SHAP), providing a more nuanced understanding of how individual features contributed to predictions [8].

Given the original model's difficulty handling outliers, particularly for *The Office,* we incorporated a winsorization step to mitigate the impact of extreme values. The final model was evaluated using an 80/20 train-test split at the TV show level.

Beyond regression, we also implemented a similarity scoring system. This approach aggregated NLP features across episodes for each show and calculated Euclidean distances to identify other shows with similar sentiment profiles. We hypothesized that this could surface shows with comparable emotional arcs in their scripts.

### D. Combined Model

After developing both the Rolling XGBoost model and the feature selection model, we attempted to combine the two approaches by first identifying the top 10 most important features using the full model, then applying the rolling approach with only those features. However, this combined model proved less accurate than either model on its own, so we ultimately decided to maintain the two models separately.

## 5. RESULTS

Due to the varying emotional tones of individual shows and the distinct audiences that different genres attract, it was difficult to generalize results across all series using a single, unified model. Instead, we found greater success by modeling individual shows separately to assess the predictive capacity of our approach. In these show-specific models, previous viewership consistently emerged as one of the top predictors of future viewership. The frequent appearance of both prior viewership and moving average features among the most important variables suggests a strong relationship between a show's past performance and its future success.

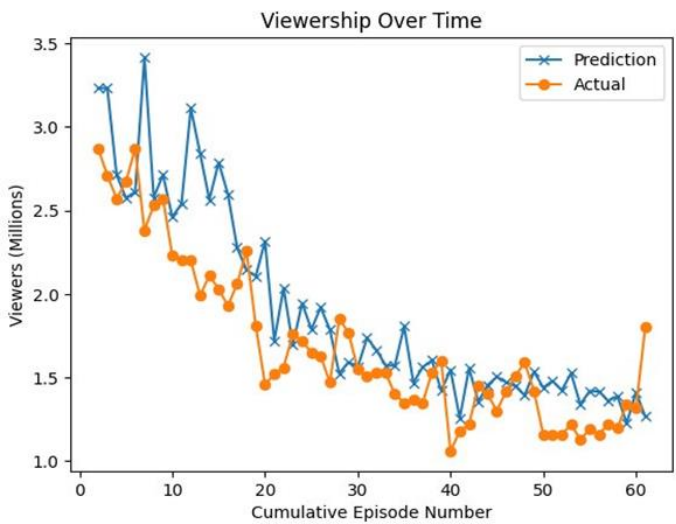

*Figure 1: Graph of predicted vs actual viewership over time of Better Call Saul using the Rolling XGBoost model.*

The Rolling XGBoost model showed varying levels of predictive accuracy across shows. For example, it performed well on *Better Call Saul*, achieving an $R^2$ of 0.742 and an RMSE of 0.361. As shown in the figure above, the model's predictions generally overestimated actual viewership, but closely tracked its overall trend across the show's run.

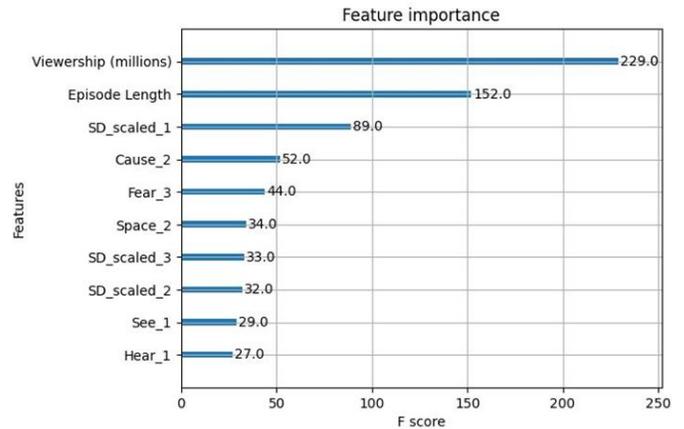

*Figure 2: Feature importance plot for Better Call Saul using the Rolling XGBoost model.*

It is also clear based on the feature importance graph that previous viewership was by far the most important predictor of future viewership. This is unsurprising because often people will come back and watch the same shows repeatedly if they've been watching them for a while, often regardless of what may happen in one episode.

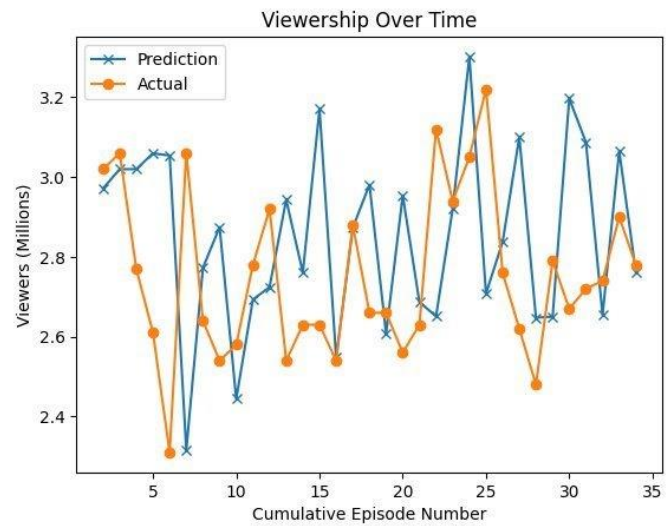

*Figure 3: Graph of predicted vs actual viewership over time of Abbott Elementary using the Rolling XGBoost model.*

In contrast to the *Better Call Saul* example, *Abbott Elementary* presented a case where the model's predictive power was minimal. Although the model achieved a relatively low RMSE of 0.330, suggesting accurate numerical predictions, the $R^2$ value was just 0.039, indicating that the model explained very little of the variance in the viewership. This apparent accuracy was largely due to the narrow range of actual viewership values, which clustered between 2.4 and 3.2 million. Interestingly, viewership was not among the top 10 most important features when the model was trained solely on *Abbott Elementary* data.

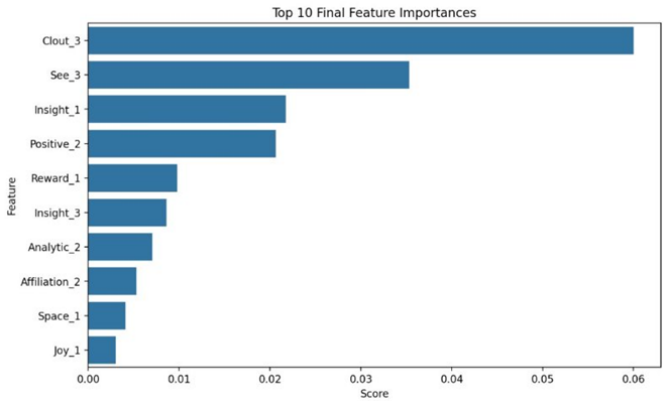

*Figure 4: Feature importance plot for Abbott Elementary using the Rolling XGBoost model.*

The feature selection model enabled us to reduce the dimensionality of the dataset and concentrate on the most influential predictors for each individual show. Beyond improving model efficiency, this approach also offers practical insights for writers and producers by highlighting which emotional scores are most strongly associated with viewership. These insights could guide content development by identifying key narrative elements that impact audience engagement. For example, when applied to *Better Call Saul*, the model achieved an RMSE of 0.208 and an $R^2$ of 0.763. The results suggest strong predictive performance and alignment with the original boosted model.

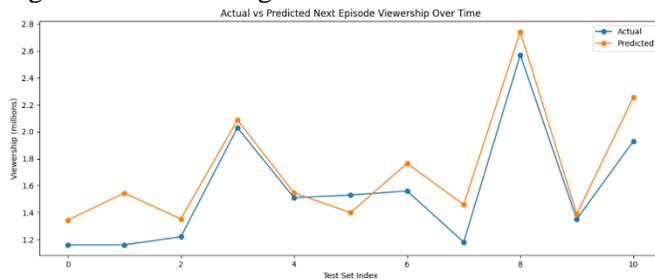

*Figure 5: Graph of predicted vs actual viewership over time of Better Call Saul using the feature selection model.*

The top three features by gain importance were previous viewership and the two moving averages of viewership with window sizes of 3 and 5 episodes. This corroborated the results of the other model and helped support the claim that previous viewership is a significant driver of next episode viewership.

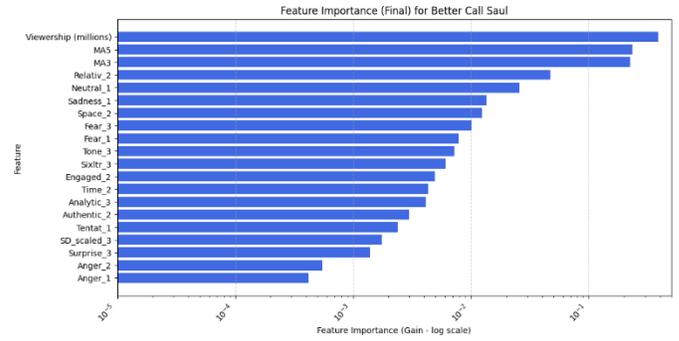

*Figure 6: Final feature importance (log-gain scale) for Better Call Saul using the feature selection model.*

*Abbott Elementary* did not perform well using this model, yielding an RMSE of 0.420 and an $R^2$ of –0.478. This negative $R^2$ indicates that the model performed worse than a baseline model that simply predicts the mean viewership for all episodes. One possible explanation is the limited test set, only 20% of the data, or seven episodes in this case, which may have made the validation less reliable compared to the rolling model's iterative testing approach. Regardless of the validation method, *Abbott Elementary* appears to be a poor fit for this modeling framework.

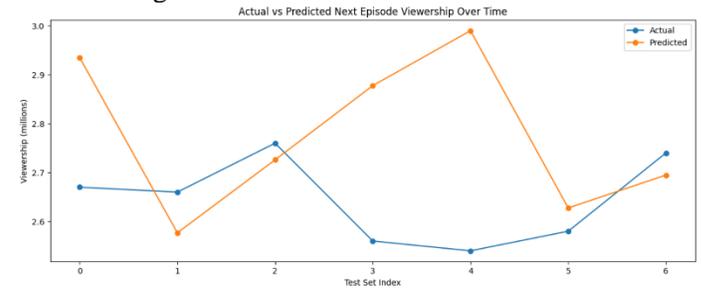

*Figure 7: Graph of predicted vs actual viewership over time of Abbott Elementary using the feature selection model.*

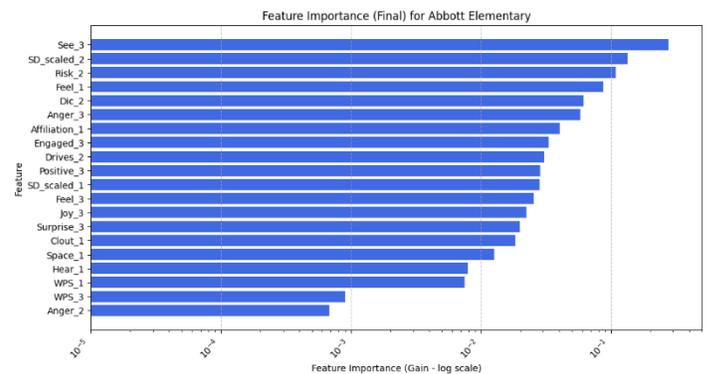

*Figure 8: Final feature importance (log-gain) scale for Abbott Elementary using the feature selection model.*

When cross-referencing Shapley (SHAP) values with gain-based feature importance, we found that the top NLP features frequently appeared in both rankings, though not always in the same order. SHAP values provide additional interpretability by indicating not only the direction (positive or negative) of a feature's effect on predictions, but also by visually encoding the magnitude of the feature's value through color [8]. This allows for a more nuanced understanding of how specific features influence viewership. For instance, in *Abbott Elementary*, the model showed a significant net negative effect on next-episode viewership when language associated with visual references or imagery in the third act appeared infrequently or had low values.

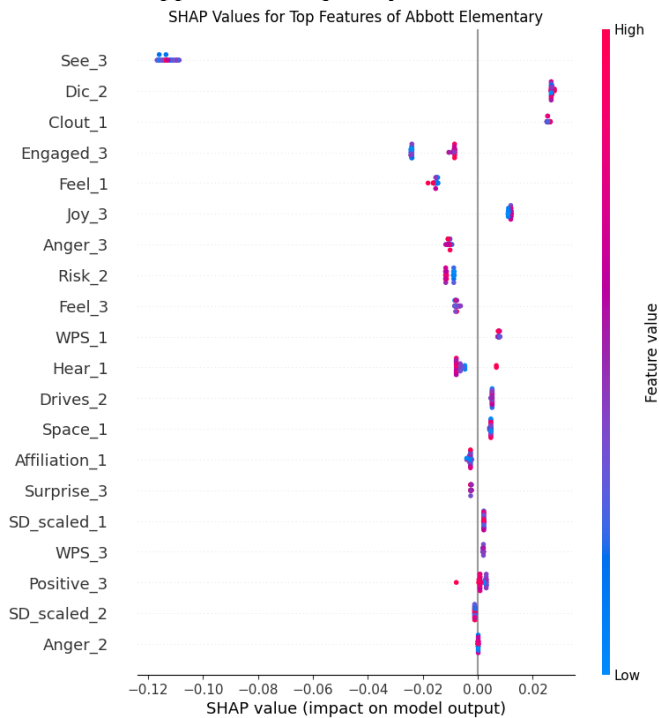

*Figure 9: SHAP values for Abbott Elementary (excluding viewership and moving averages).*

When examining the similarity scores for each show, calculated using Euclidean distance between the aggregate NLP scores, we primarily relied on human judgment to assess the plausibility of the results. For some shows, the matches were intuitive. *Breaking Bad* emerged as the most similar show to *Better Call Saul*, which made sense given that it is the original series from which *Better Call Saul* was spun off. Similarly, the top result for *CSI: Miami* was *CSI: Crime Scene Investigation*, a logical pairing since both were created by the same production company and share a common concept. However, other results were more difficult to interpret. For example, in the case of *The Shield*, the top five similar shows unexpectedly included *Regular Show*, *Adventure Time*, and *Rick and Morty*, a surprising outcome given the stark differences in genre, tone, and target audience.

## DISCUSSION

When comparing RMSE plots for both the rolling XGBoost model and the feature selection model, we found that both approaches frequently achieved an RMSE of under one million viewers. However, the rolling model exhibited a slightly longer tail of higher RMSE values, suggesting that the feature selection model was marginally more consistent in its predictions. To investigate the factors influencing $R^2$ values, we plotted distributions by genre for both models and found no meaningful correlation between genre and $R^2$ performance.

Interpreting feature importance using XGBoost's built-in gain metric was useful in identifying the degree to which the model relied on autoregressive features. However, SHAP values provided a more detailed understanding of how specific features influenced viewership, enabling a clearer interpretation of the direction and magnitude of each feature's effect. This aligns more closely with the goals of our project by helping to pinpoint which script elements contribute to increases or decreases in viewership.

When examining the results of our similarity function, based on Euclidean distances between NLP score vectors, we were surprised by some of the top-ranked similar shows. While some results defied expectations, we argue that this method may surface latent similarities not readily apparent to industry professionals. As such, it offers value by enabling writing teams to contextualize their work in relation to other series, justify creative decisions, and benchmark against prior industry content.

## CONCLUSION

It may not come as a surprise to those in the entertainment industry that the previous episode viewership is often the strongest predictor of the next episode viewership. However, this modeling approach offers added value by revealing which features are most influential in specific formats, such as *The Office* and *Abbott Elementary*. This can guide writers aiming to craft the next successful mockumentary by highlighting the narrative elements that have historically resonated with audiences. Moreover, the metrics generated allow for meaningful comparisons across shows, providing network executives with a richer historical context and enabling more informed predictions about the potential success of pilots and currently airing series.

## Acknowledgments

The authors would like to express their sincere gratitude to Anthony Palomba for bringing this project idea forward and for his invaluable insights, feedback, and industry expertise that shaped the direction of this research. We also thank Tom Hartvigsen for his dedicated mentorship and guidance throughout the project, providing critical support in both technical and strategic aspects of our work.